\documentstyle[12pt]{article}
\begin{document}

\begin{center}
\begin{large}
{\bf Hartree-Fock ground-state energy of anyons\\
with no Coulomb interaction in the zero effective field}
\end{large}
\end{center}

\vspace{1cm}

\noindent Piotr Sitko, Institute of Physics,
Technical University of Wroc\l{}aw,\\ Wyb. Wyspia\'{n}skiego
 27, 50-370 Wroc\l{}aw, Poland.

 \vspace{1cm}
\begin{center}

Abstract

\end{center}
We find, in the Hartree-Fock approximation, the ground-state
energy of anyons with no Coulomb interaction in the case
when the external magnetic field
precisely cancels the average statistical field. From the point of
view of the fractional quantum Hall effect it is shown that
statistics transmutations to {\it
superfermions} at filling fractions $\nu =1/2p$ are not energetically
favourable.

\vspace{2cm}
\newpage
\noindent{\bf  Introduction}

It was shown by Greiter and
Wilczek \cite{Greiter}, and Jain \cite{Jain}, that
the fractional quantum Hall effect (FQHE)  can be
explained if one considers a statistics transmutation to {\it
superfermions}. It means that an even number ($2p$) of flux quanta is
attached to each electron (such a construction does not change the
antisymmetrity of the wave function). Averaging fictious fluxes one
finds the mean statistical field
$B^{s}=\Phi^{s}/S= \rho\cdot 2p\frac{hc}{e}$
($S$ - the area of the system)
which partly cancels the external magnetic field $B^{ex}$, e.i.
the effective field acting on electrons is
$B^{ef}=B^{ex}-B^{s}$. Hence, the  filling fractions at which one
obtains completely filled Landau levels are given by $\nu =
n/(2pn\pm 1)$ ($n$ - an integer number). The Hartree-Fock
approximation (HFA)
of this system \cite{Cabo} confirms the presence of Greiter-Wilczek
states.

Recently Halperin, Lee and Read \cite{Halperin} have proposed that at
filling fractions with even denominator one also deals with the
statistics transmutation with attaching $2p$ flux quanta to each
electron. This is especially important at the fractions $\nu
=(2p)^{-1}$	when there is a precise cancelation of  magnetic and
 statistical fields, i.e. $B^{ef}=0$. In this case the ground state
of the system is belived to be a Fermi sea with the energy
$N\frac{E_{F}}{2}$ while the energy of electrons in the
magnetic field $B^{s}=B^{ex}$	is $N\frac{E_{F}}{2}2p$. Hence, for a
first glimpse, one finds a statistics transmutation energetically
favourable at $\nu =(2p)^{-1}$.

The RPA description presented {\it et al.}\cite{Halperin} includes
only  effective two-body interactions \cite{Chen,Dai} and omits
other interaction terms. In the case of anyons it is the HFA which
involves the full interaction Hamiltonian
\cite{Hanna,Fetter,Sitko}.
The aim of this paper is  to verify in the HFA the
ground state and the energy of the system of electrons in the magnetic
field at	filling fraction
$\nu =(2p)^{-1}$
while attaching $2p$ flux quanta
to each electron.
\newpage
\noindent{\bf  Hartree-Fock ground-state energy}

Let us start from the Hamiltonian of anyons (in the fermion
representation) in the external magnetic
field \cite{Chen}:
\begin{equation}
H=\frac{1}{2m}\sum_{i=1}^{N}({\bf P}_{i}+\frac{e}{c}{\bf
A}_{i}-\frac{e}{c}{\bf
A}_{i}^{ex})^{2}
\end{equation}
where
\begin{equation}
{\bf A}_{i}=\frac{2p \hbar c}{e}	\hat{\bf z}\times \sum_{j\neq i}
\frac{({\bf r}_{i}-{\bf r}_{j})}{|{\bf r}_{i}-{\bf r}_{j}|^{2}}\;\; ,
\end{equation}
$\hat{\bf z}$ - a perpendicular to the plane unit vector.
The average statistical  field vector potential is given by the equation:
\begin{equation}
\bar{\bf A}_{i}=\frac{2p \hbar c \rho}{e}\hat{\bf z}\times\int
\frac{({\bf r}_{i}-{\bf r}_{j})}{|{\bf r}_{i}-{\bf r}_{j}|^{2}}
d^{2}{\bf r}=\frac{1}{2}B^{s}\hat{\bf z}\times {\bf r}_{i} \; .
\end{equation}
In this paper we assume that $B^{s}=B^{ex}=\nabla\times {\bf
A}^{ex}$.
Let us note that
the integral (3) is formally divergent, however, for finite areas it gives
the right average field $B^{s}=\nabla\times \bar{\bf A}$	\cite{Jacak}.
The above result
is obtained in a circle.

The Hamiltonian (1) for  prominent cases, $p$ - being an integer,
describes the statistics transmutations to {\it superfermions} and for
other values of $p$ - the transmutations to anyons with  statistics
parameters
$\Theta = (1-2p)\pi$ (an exchange of two anyons produces a phase
factor of $e^{i\Theta}$).

Dividing the Hamiltonian (1) into one-, two- and three-body parts one
finds the Hartree-Fock equations of the ground-state energy
\cite{Hanna,Fetter,Sitko}:
\begin{equation}
<H_{a}>=\frac{1}{2m}\sum_{\bf p}\int d1 \phi_{\bf p}^{+}(1){\bf
P}_{1}^{2}\phi_{\bf p}(1)\;\; ,
\end{equation}
\begin{equation}
<H_{b}>=-\frac{e}{mc}\sum_{{\bf p},{\bf q}}^{\:}^{'} \int d1d2\phi_{\bf
p}^{+}(1)\phi_{\bf q}^{+}(2){\bf A}_{12}{\bf P}_{1}\phi_{\bf p} (2)
\phi_{\bf q} (1)\; ,
\end{equation}
\begin{equation}
<H_{c}>=\frac{e^{2}}{2c^{2}m}\sum_{{\bf p},{\bf q}}^{\;}^{'}
\int d1d2\phi_{\bf
p}^{+}(1)\phi_{\bf q}^{+}(2)|{\bf A}_{12}|^{2}\phi_{\bf p}(1)
\phi_{\bf q}(2)\; ,
\end{equation}
\begin{equation}
<H_{d}>=-\frac{e^{2}}{2c^{2}m}\sum_{{\bf p},{\bf q}}^{\;}^{'}
\int d1d2\phi_{\bf
p}^{+}(1)\phi_{\bf q}^{+}(2)|{\bf A}_{12}|^{2}\phi_{\bf p}(2)
\phi_{\bf q}(1)\; ,
\end{equation}
\begin{equation}
<H_{e}>=\frac{e^{2}}{c^{2}m}\sum_{{\bf p},{\bf q},{\bf k}}^{\;}^{''}\int
d1d2d3
\phi_{\bf p}^{+}(1)\phi_{\bf q}^{+}(2)\phi_{\bf k}^{+}(3)
{\bf A}_{12}{\bf A}_{13}
\phi_{\bf p}(3)\phi_{\bf q}(1)\phi_{\bf k}(2)\; ,
\end{equation}
\begin{equation}
<H_{f}>=-\frac{e^{2}}{2c^{2}m}\sum_{{\bf p},{\bf q},{\bf k}}^{\;}^{''}\int
d1d2d3
|\phi_{\bf p}(1)|^{2}\phi_{\bf q}^{+}(2)\phi_{\bf k}^{+}(3)
{\bf A}_{12}{\bf A}_{13}
\phi_{\bf q}(3)\phi_{\bf k}(2)
\end{equation}
where $d1=d^{2}{\bf r}_{1}$, sums extend over the Fermi sea and
omit  elements of two equal momentum variables. We assume the wave
function to be
the plane wave \\ $\phi_{\bf
p}(1)=S^{-\frac{1}{2}}e^{\frac{i}{\hbar}{\bf p}{\bf r}_{1}}$.

One can use a Fourier transform of a 2D Coulomb potential to write
\cite{Dai}:
\begin{equation}
\int\frac{\bf r}{|{\bf r}|^{2}}e^{\frac{i}{\hbar}{\bf p}{\bf r}}
d^{2}{\bf r}=
\frac{2\pi\hbar}{i}\frac{\bf p}{|{\bf p}|^{2}}
\end{equation}
and the fact that in 2D a Fermi sea is a circle to obtain (using
Eq.(3)):
\begin{equation}
\sum_{{\bf p}'\neq {\bf p}\; ,|{\bf p}'|<p_{F}}
\frac{({\bf p}-{\bf p}')}{|{\bf p}-{\bf p}'|^{2}}=
\pi \rho_{p}{\bf p}
\end{equation}
where $\rho_{p}=\frac{S}{(2\pi\hbar)^{2}}$ - the  density of states in the
momentum space.
Using relations (10) and (11) one finds:
\begin{equation}
<H_{a}>=N\frac{E_{F}}{2},\;\;\; <H_{b}>=0,
\end{equation}
\begin{equation}
<H_{c}>=\frac{E_{F}p^{2}}{\pi}\rho
\int d1d2\frac{1}{|{\bf r}_{1}-{\bf r}_{2}|^{2}}
\end{equation}
and
\begin{equation}
<H_{d}>=p^{2}N\frac{E_{F}}{2},\;\; <H_{e}>=2p^{2}N\frac{E_{F}}{2},
\end{equation}
\begin{equation}
<H_{f}>=-\frac{E_{F}p^{2}}{\pi}\rho_{p}
\int d^{2}{\bf k}d^{2}{\bf q}\frac{1}{|{\bf k}-{\bf q}|^{2}} \;\; .
\end{equation}
Since $<H_{c}>+<H_{f}>=0$ the final result is finite:
\begin{equation}
<H>=N\frac{E_{F}}{2}(1+3p^{2})
\end{equation}
which   confirms the choice of the ground state.

Let us consider the transmutation to bosons, e.i. $\Theta =0$
($p=\frac{1}{2}$). The obtained energy  is $\frac{7}{4}$ of the
energy of bosons in the magnetic field $B^{ex}=2p
\frac{hc\rho}{e}=B^{s}$. Such a discrepancy appears also in the HFA
of anyons \cite{Hanna} and it should become negligible in a higher
order approximation \cite{Fetter}.

In the case of $p$ being an integer the  interaction energy
is many times ($3p^{2}$) higher than the energy of free electrons
which shows how strong statistical interactions of {\it superfermions}
are.
If one takes the system of electrons in the external magnetic field
$B^{ex}=B^{s}$	the cost of the statistics transmutation to {\it
superfermions} is
$\Delta E=N\frac{E_{F}}{2}(2p^{2}+(p-1)^{2})>0$ and increases rapidly
with $p$.

\noindent{\bf  Conclusions}

We have confirmed, in the Hartree-Fock approximation, that the ground
state of anyons in the zero effective field is the Fermi sea.
Moreover, the ground-state energy is $N\frac{E_{F}}{2}(1+3p^{2}) $
(the statistics parameter $\Theta =(1-2p)\pi$).
Comparing with the energy of electrons in the magnetic field
$B^{ex}=B^{s}$, $N\frac{E_{F}}{2}2p$, we conclude that in the FQHE
the generation
of $2p$ flux quanta attached to each electron is not energetically
favourable at filling fraction $\nu=(2p)^{-1}$. The above analysis
(HFA) needs to be verify in a higher order approximation e.g. the RPA
with the full interaction Hamiltonian \cite{Dai}.

\noindent{\bf  Acknowledgements}

I would like to thank Lucjan Jacak, Andrzej Radosz and Liliana
Bujkiewicz for helpful discussions.

\end{document}